\documentclass[aps,pra,reprint,groupedaddress,showpacs]{revtex4-1}

\usepackage{amsmath,amssymb,graphicx}

\begin{document}
\newcommand{\LP}[2]{\ensuremath{\text{LP}_{#1#2}}}%
\newcommand{\LPx}[2]{\ensuremath{\text{LP}_{#1#2,\text{x}}}}%
\newcommand{\LPy}[2]{\ensuremath{\text{LP}_{#1#2,\text{y}}}}%

\title{Optically induced mode conversion in graded-index fibers using ultra-short laser pulses}

\author{Tim Hellwig}
\email{Corresponding author: tim.hellwig@uni-muenster.de}
\author{Till Walbaum}
\author{Carsten Fallnich}
\affiliation{Institute of Applied Physics, Westf\"alische Wilhelms-Universit\"at, 48149 M\"unster, Germany}
\date{\today}

\begin{abstract} 
We propose the use of graded-index few-mode fibers for mode-conversion by long-period gratings (LPG)
transiently written by ultrashort laser pulses using the optical Kerr effect. 
The mode interaction is studied by numerically solving the multi-mode coupled nonlinear Schr\"odinger equations.
We present highly efficient conversion of the \LP01- into the \LP11-mode preserving the pulse shape in contrast to previous results in step-index fibers. Furthermore, mode conversion using different wavelengths for inducing and probing the LPG is shown. Due to the flat phase-matching curve of the examined modes in the graded-index fiber, mode-conversion can be observed for probe center wavelengths of 1100\,nm up to 1800\,nm with a write beam centered around 1030\,nm. Therefore, a complete separation of the probe from the write beam should be possible as well as the application of optically induced guided mode conversion for all optical modulation across a broad wavelength range.
\end{abstract}


\maketitle 

\section{Introduction}
Higher-order guided transverse modes in few-mode fibers have drawn a lot of attention recently due to their various applications, e.g., in bandwidth enhancement for optical communications \cite{Ryf2012} including few-mode optical amplification \cite{LeCocq2012}, dispersion compensation  \cite{Ramachandran2005}, or optical switching \cite{Nguyen2008}.  Especially in the context of dispersion compensation guided mode conversion using long-period gratings (LPG) has been studied extensively with the focus on permanent gratings written by UV light in photosensitive fibers or by femtosecond pulses \cite{Kondo1999}. 

An alternative approach by Andermahr et al. \cite{Andermahr2010} is based on a transient LPG written via the optical Kerr effect. The LPG was generated by launching a high power write beam into a mixture of two transverse modes \LP01 and \LP11. Their difference in propagation constant $\Delta\beta^{\text{write}}=\beta_{\text{LP}_{01}}-\beta_{\LP11}$ leads to the formation of a spatially inhomogeneous but periodic intensity pattern in the fiber due to multimode interference.  The intensity pattern is directly translated into a refractive index modulation via the optical Kerr effect. Using this induced LPG, mode conversion could be achieved by converting a counterpropagating probe beam from the \LP01- to the \LP11-mode. A conversion efficiency of about 50\% could be demonstrated experimentally, limited by the damage threshold of the fiber front facet at the used pulse energies. In order to reduce the required pulse energy for the write beam Walbaum et al. proposed the use of femtosecond laser pulses for writing and probing the LPG in an extensive numerical study \cite{Walbaum2012}. With femtosecond pulses, peak intensities beyond $100$\, kW are achievable well below the damage threshold of fiber facets and in a first experimental verification guided mode conversion from the \LP01- to the \LP11-mode as well as to the \LP02-mode by an optically induced LPG could be demonstrated \cite{Walbaum2012a} with pulse energies reduced by a factor of 300.

So far the investigated ultra-fast mode conversion approaches are still restricted by the following issues:
On the one hand (1) the maximum achievable conversion efficiency and the quality of the temporal shape of the converted pulse are severely limited due to  (i) mode walk-off between the write pulses, (ii) the phase-matching bandwidth of the induced LPG \cite{Schaferling2011} as well as a (iii) temporal varying LPG strength according to the write pulse's temporal shape.
On the other hand (2) experimental difficulties arise when separating probe and write beam by polarization due to residual birefringence in the fiber as well as nonlinear polarization rotation for a non-zero phase-difference between both beams \cite{Walbaum2012a}.

In this paper we show that graded-index fibers are a promising alternative to step-index fibers to overcome the aforementioned problems. We numerically study guided mode conversion in graded-index fibers by solving the coupled nonlinear Schroedinger equations for each mode \cite{Poletti2008}. In the first part of the paper (section~\ref{Sec:EquWav}) we focus on issue (1): in a scenario similar to the one presented in \cite{Walbaum2012} we study the general difficulties achieving a high conversion efficiency as well as preserving the temporal pulse shape. We specifically address how graded-index fibers allow to reduce the limitations posed by (i) to (iii) when working with ultrashort pulses for mode conversion. Furthermore, in the second part (section~\ref{Sec:Dichroic}) we demonstrate how, due to the broad bandwidth of the phase matching condition in graded-index fibers, a detuning of the probe beam over a wide spectral range is made possible. The resulting two-color mode conversion scenario will allow to separate the probe beam from the write beam using, e.g., dichroic filters and thereby to circumvent the experimental challenges (2) arising from having to discriminate probe and write beam by polarization. Finally, in the last part (section~\ref{sec:Modu}) a possible application of mode conversion with preserved temporal pulse shape for high quality all-optical modulation is presented.
  
\section{Mode conversion with identical write and probe center wavelengths}
\label{Sec:EquWav}

We have chosen a typical graded-index fiber for our studies with a $50\,\mu$m core and a parabolic refractive index profile with a peak refractive index difference of $\Delta n=0.0138$ at 1300\,nm, thereby falling under the specifications of the ITU-T G.651.1 recommendations \cite{G651}.
The propagation constants and mode fields of the $\text{LP}_{\text{lm}}$-modes are calculated by numerically solving a matrix eigenvalue problem derived from the scalar Helmholtz equation with the finite-difference approximation \cite{Riishede2003}. The mode propagation constants $\beta_{lm}(\omega)$ are obtained in dependence of the frequency and then approximated by a Taylor expansion up to the fifth order. These dispersion constants and the mode fields are used to numerically solve the multi-mode nonlinear Schr\"odinger equation presented by Poletti et al. \cite{Poletti2008} using a split-step Fourier algorithm.

Similar to our earlier work in step-index fibers \cite{Walbaum2012} we start our considerations with two co-propagating, cross-polarized beams both at the center wavelength of $\lambda_0=1030\,$nm. All laser pulses in this work are considered to have an initially Gaussian temporal shape. The y-polarized beam constitutes the write beam and is launched into the fiber in an equal mixture of the \LP01- and the \LP11-mode. These two mode components induce a long-period grating due to their mode interference and the intensity-dependent refractive index.
When working with ultrashort pulses, the mode conversion efficiency, meaning the fraction of power that can be transferred to another mode, depends not only on the phase mismatch 
\begin{equation} \Delta= \Delta\beta^{\text{write}}(\lambda_\text{write})-\Delta\beta^{\text{probe}}(\lambda_\text{probe}),
\end{equation}
but also on the temporal walk-off 
\begin{equation} \Delta\beta_1= \beta_{1,\text{LP}_{lm}}(\lambda)-\beta_{1,\text{LP}_{l'm'}}(\lambda')
\end{equation}
 between the individual pulses propagating along the fiber, in the case considered in this section exclusively due to mode walk-off as the x-polarized probe beam is at the same center wavelength as the write beam.  Therefore, the conversion from the \LPx01-mode to the \LPx11-mode is also inherently phase-matched ($\Delta=0$) to the LPG written by the write modes. The mode `conversion speed' \cite{Walbaum2012}, meaning the propagation length necessary to reach the maximum conversion, depends on the coupling strength of the grating and thereby on the time-dependent intensity of the write beam, leading to a temporal non-uniform conversion of the probe pulse.   

\begin{figure}[htbp]
\includegraphics[width=1\columnwidth]{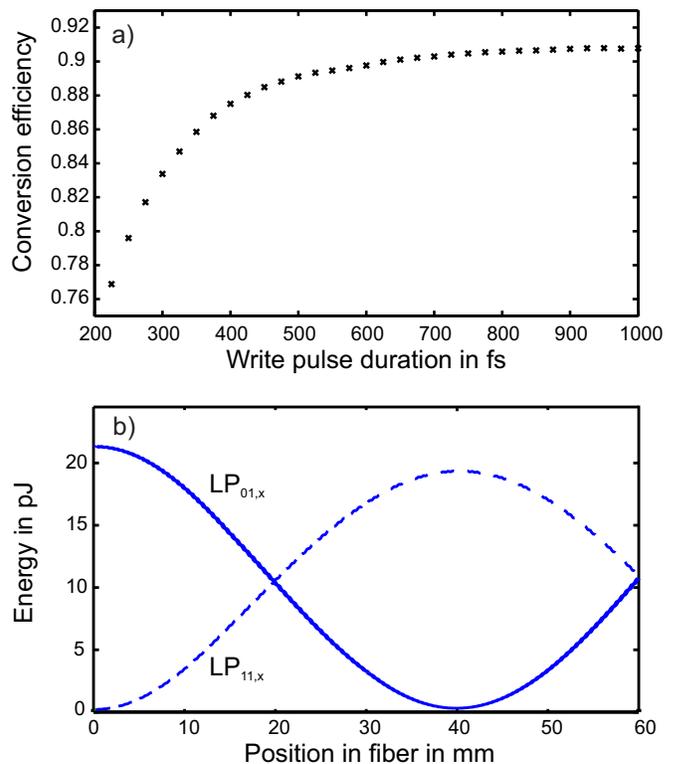}
\caption{\label{Fig:1}Mode conversion of a 200\,fs probe pulse in a graded-index fiber (core diameter of  $50\,\mu$m), when write and probe beam have the same wavelength. a) Conversion efficiency in dependence of the write beam's pulse duration and b) detailed evolution of the probe beam's mode content along the fiber, when the LPG is induced by a 600\,fs long write pulse.}
\end{figure}

Furthermore, graded-index fibers are specially designed to reduce mode walk-off, resulting in a very small group delay difference of $\Delta\beta_{1} = 0.16\,\tfrac{\text{fs}}{\text{cm}}$ between the two modes investigated here, compared to standard step index fibers ($\Delta\beta_{1} \approx 6 - 27\,\tfrac{\text{fs}}{\text{cm}}$ in \cite{Walbaum2012}). This leads to a negligible mode walk-off for the fiber lengths considered here. To also reduce the temporal non-uniform conversion due to the temporal profile of the refractive index modulation and thereby the LPG, we studied the influence of the write beam's pulse duration on the conversion efficiency (see Fig.~\ref{Fig:1}a). The probe beam's pulse parameters were fixed at 200\,fs pulse duration and 100\,W peak power while we varied the write beam's pulse duration from 200\,fs to 1\,ps with a constant peak power of 60\,kW in each mode. In Fig. \ref{Fig:1}a an increase in conversion efficiency  with pulse duration can be seen that saturates at longer pulse durations. This is due to the more homogenous conversion in the time domain for longer write pulses which reduces backconversion at the pulse peak when the pulse wings have not yet been converted to the higher-order mode.
Efficient conversion with about 90\% of the inital probe energy transfered to the higher-order mode is found for write pulse durations of longer than 600\,fs. 

\begin{figure*}[htbp]
\includegraphics[width=2\columnwidth]{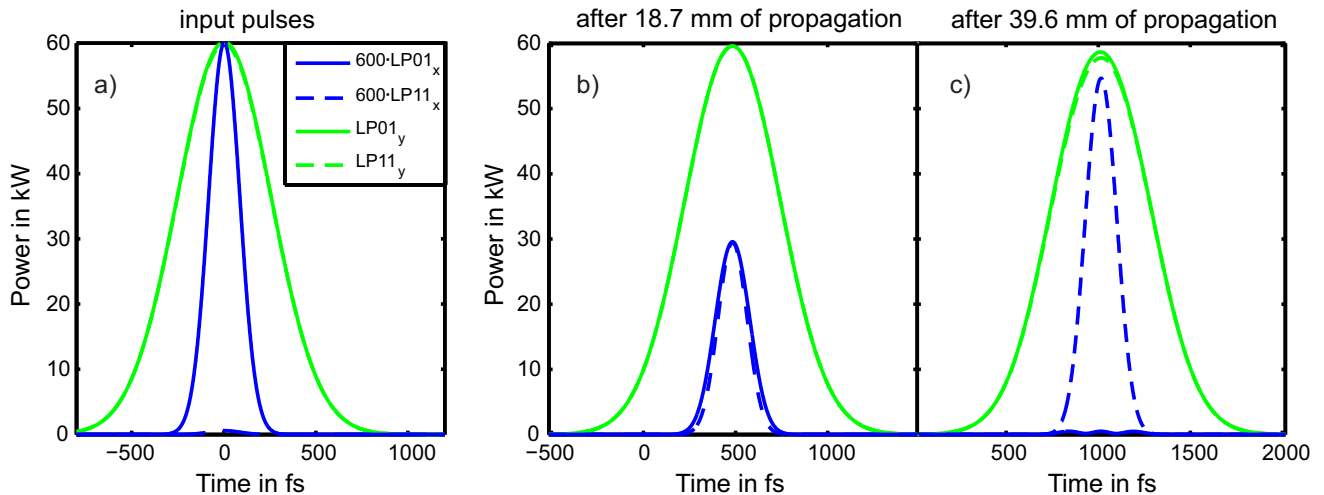}
\caption{\label{Fig:2}Mode composition of the probe and write pulse during propagation along the $50\,\mu$m core diameter graded-index fiber. Pump and probe pulse have the same center wavelength and are considered as cross-polarized at the beginning of the fiber (a). Depicted are the points of (b) equal peak power and (c) maximum conversion of the two modes of the probe pulse. The probe modes are scaled in all sub-figures by a factor of 600 for better visibility.}
\end{figure*}

The detailed result of the energy transfer between the two probe modes for 600\,fs long write pulses is displayed in Fig.~\ref{Fig:1}b, where the energy content of the two modes during propagation along the $50\,\mu$m core diameter graded-index fiber is shown. After about $2$\,cm of propagation the two modes contain equal energy, and after about $4$\,cm maximum conversion is achieved with 90\% of the input energy converted to the \LP11-mode, and the \LP01-mode is still only holding $1.5$\% of residual energy.
Some energy in the probe modes (in this case 8.5\%) is lost which can be attributed to four-wave mixing with the self-phase modulated spectrum of the write modes. It is verified that the overall energy in the system is conserved.

The detailed evolution of the probe and write pulses in the different fiber modes is depicted in Fig.~\ref{Fig:2}. Herein, the temporal pulse profile is decomposed into the different mode contents for the write beam (\LPy01 and \LPy11) as well as for the probe beam (\LPx01 and \LPx11). The power of the probe modes is magnified by a factor of $600$ for better visibility. In Fig.~\ref{Fig:2}a the input modes are shown while Fig.~\ref{Fig:2}b and Fig.~\ref{Fig:2}c depict the mode contents after $18.7$\,mm and $39.6$\,mm of fiber propagation, at the points of equal peak power and of maximum mode conversion, respectively.
At the beginning of the fiber (Fig.~\ref{Fig:2}a) the probe beam is launched into the \LPx01-mode of the fiber with 200\,fs pulse duration which is by a factor of 3 shorter than the write pulse. The coupling strength between the probe modes depends on the strength of the written LPG at the respective temporal position. Therefore, the longer write beam ensures that even for the wings of the probe pulse a sufficient coupling strength is still present. 

After $18.7$\,mm of propagation the probe pulse is decomposed into two modes with equal peak power. As it can be seen from Fig.~\ref{Fig:2}b, the Gaussian temporal shape of the remaining energy in the \LP01-mode is well preserved and nearly identical to the temporal shape of the converted \LP11-mode. The part of the probe pulse, that converted to the \LP11-mode, also has a Gaussian shape with a slightly shorter pulse duration ($185.5$\,fs for the \LP11-mode compared to $215.8$\,fs for the \LP01-mode). The difference in pulse duration can be explained by the temporal shape of the write pulse inducing the LPG. The conversion speed is determined by the amount of refractive index change. The temporal width of the electrical field of the converted probe pulse can therefore be approximated as the product of the Gaussian shaped refractive index modulation (with a temporal width corresponding to the FWHM intensity of the write pulse) and the temporal width of the electrical field of the original probe pulse. Following this simple model the FWHM of the converted part of the probe pulse (\LPx11) is calculated to $185.1$\,fs from the final \LPx01 pulse duration of $215.8$\,fs. Although the continuous broadening of the probe and write pulse due to dispersion is neglected in this simple model the result compares to the simulated data very well. 

Finally, after $39.6$\,mm of propagation, almost the entire probe pulse energy (1.5\% residual energy left in \LPx01) has been converted from the \LPx01- to the \LPx11-mode. The pulse shape is still fitted very well by a Gaussian shape and has a pulse duration of $199$\,fs. Compared to our previous work this is made possible by the small group delay difference on the one hand and by the flat phase matching curve of the conversion in the graded-index fiber on the other hand. At a fiber length of about $4\,$cm the walk-off between the observed modes is still below $1\,$fs while it would have been between about $24$ and $108\,$fs in the step-index fibers investigated in \cite{Walbaum2012}. Furthermore, the phase mismatch $\Delta (\lambda)$ is below $0.008\cdot\tfrac{2\pi}{\text{cm}}$ in a broad ($>100\,$nm) window around the center wavelength. This allows a uniform conversion in the spectral domain even though a spectral broadening and modulation of the write pulse due to self-phase modulation is observed in the simulation. 

Overall, our result demonstrates the optically induced, complete conversion from the fundamental mode of the fiber to the first higher-order mode with preserved pulse shape and duration. It was verified that no mode conversion can be observed without a write beam present.
 
\section{Dual-color mode conversion}
\label{Sec:Dichroic}

The simulated mode conversion presented above - even though it is a significant improvement compared to our previous work - still includes the same experimental challenges that were encountered in \cite{Walbaum2012a} and theoretically studied in \cite{Walbaum2012}: Due to the distinction of probe and write beam by polarization, nonlinear polarization rotation as well as crosstalk due to linear birefringence can easily mask the mode conversion in the experiment. Therefore, a dual-color approach to the mode conversion seems to be promising as combining and separating the probe and write beam can then be easily accomplished by using, e.g., dichroic filters regardless of the polarization.

For different probe and write wavelengths the conversion efficiency still depends on the phase matching $\Delta (\lambda)$ as well as the occurring walk-off. The origin of the walk-off is now no longer limited to mode group delay $(\beta_{1,\LP01}\neq\beta_{1,\LP11})$, but also includes chromatic group delay $(\beta_{1}(\lambda_\text{write})\neq\beta_{1}(\lambda_\text{probe}))$.
Ideally the phase mismatch  between the write and probe beam has to be zero in order to maximize the conversion efficiency \cite{Schaferling2011}. For different probe and write wavelengths, but considering the same mode pair for both beams (the \LP01- and \LP11-mode), $\Delta(\lambda)=0$ can not be fulfilled in the graded-index fiber investigated here. Nevertheless, the phase matching curve is very flat: the phase mismatch increases approximately linearly from zero at a probe wavelength equal to the write wavelength ($1030\,$nm) up to $0.04\cdot\frac{2\pi}{\text{cm}}$ at a probe wavelength of 1800\,nm, indicating that indeed mode-conversion should be possible over a broad spectral range.

\begin{figure}[tbhp]
\centerline{\includegraphics[width=1\columnwidth]{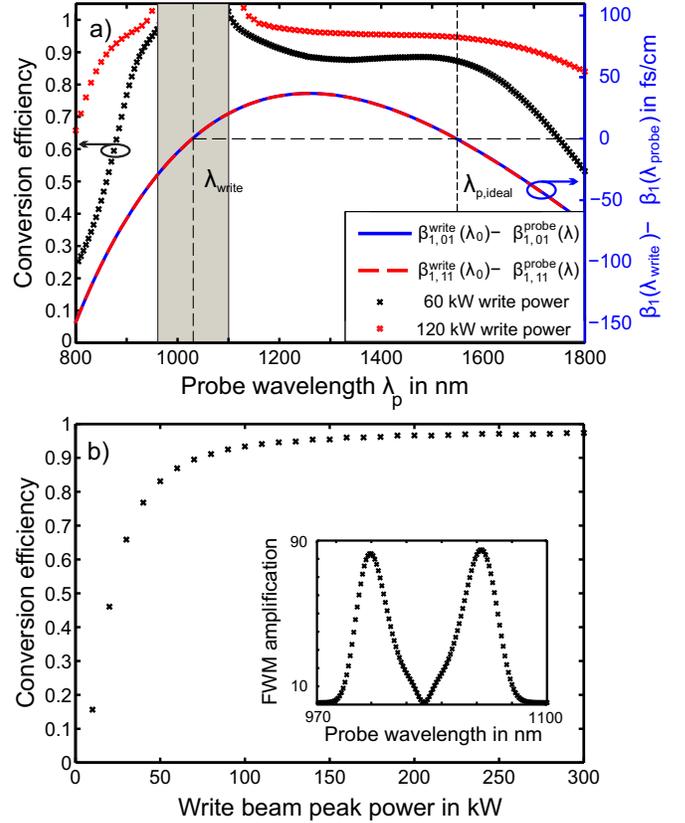}}
\caption{\label{Fig:5}Dual-color mode conversion for a write beam center wavelength of $1030\,$nm. The probe beam's center wavelength is tuned from $800$\,nm to $1800$\,nm. In (a) the wavelength-dependent conversion efficiency is plotted for two different peak powers of the write modes (60\,kW and 120\,kW) as well as the group delay difference $\beta_{1,lm}(\lambda_\text{write})-\beta_{1,lm}(\lambda_\text{probe})$ of the involved $\text{LP}_{lm}$ modes is presented. The dependence of the conversion efficiency on the write beam's peak power is investigated in detail in b) for a probe wavelength of 1550\,nm. Furthermore, in the inset in b) the occurring direct four-wave mixing cross-talk between write modes (60\,kW peak power) and probe modes in the grey-shaded area from a) is shown by plotting the total energy in probe modes after 100\,mm of propagation divided by the total input energy in the probe modes.}
\end{figure}

We studied the wavelength dependence of the mode conversion by scanning the probe beam's center wavelength for two different peak powers (60\,kW as well as 120\,kW) in each write mode, while keeping a fixed write beam center wavelength of $\lambda_\text{write} = 1030\,$nm. All other pulse and fiber parameters were kept the same as in section~\ref{Sec:EquWav}, although dual-color mode conversion would also allow to use co-polarized write and probe beams. It was verified that in a co-polarized configuration mode-conversion does occur, but in the following we will focus on the cross-polarized dual-color scenario:  In Fig.~\ref{Fig:5}a the observed conversion efficiency is plotted versus the probe beam's center wavelength. As it can be already seen for write powers of 60\,kW per write mode (black crosses in Fig.~\ref{Fig:5}a), mode conversion is achieved across a very broad spectral range (1100\,nm up to probe wavelengths of 1800\,nm) with more than 50\% efficiency. The probe wavelength in the gray-shaded area are rendered useless for mode conversion scenarios as here direct four-wave mixing redistributes energy from the write beam to the probe modes: This was studied in more detail by fine scanning the probe beams's center wavelength in the area of interest. Typically, for probe wavelengths outside the grey-shaded area, the conversion length, meaning the propagation length necessary to reach maximum conversion, varied from 60\,mm up to about 95\,mm. Therefore, the upper boundary of the direct four-wave mixing contribution was estimated by dividing the total output energy in both probe modes by the initial probe pulse energy after a propagation of 100\,mm. The result is shown in the inset in Fig.~\ref{Fig:5}b, where it becomes apparent, that the four-wave mixing amplification exceeds the mode-conversion by almost two orders of magnitude; 
To ensure that the direct four-wave mixing contribution has only a negligible influence on the mode-conversion, the wavelength region where the four-wave mixing amplification is bigger than 5\% is left out of further consideration (970\,nm up to 1100\,nm).


In order to get a detailed understanding of the conversion efficiency it is helpful to take a closer look at the involved propagation constants: hence, the group delay difference $\Delta\beta_{1,lm}(\lambda)=\beta_{1,lm}(\lambda_\text{write})-\beta_{1,lm}(\lambda_\text{probe})$ is displayed in Fig.~\ref{Fig:5}a for both involved $\text{LP}_{lm}$ modes. As it can be seen, the curves of the \LP01- and the \LP11-mode are not distinguishable in this representation as it was expected from the low mode group delay difference in graded-index fibers. For the dual-color conversion scenario this means that there is almost no modal group delay difference and thereby only negligible walk-off \textit{within} each beam. Nevertheless, as it was mentioned before, the chromatic group delay difference has to be considered as well, leading to a walk-off \textit{between} the probe and the write beam related to different wavelengths.

Two probe wavelengths are worth mentioning concerning their chromatic group delay: On the one hand, at about $1250\,$nm the group delay has a local maximum with a value of $36\,\tfrac{\text{fs}}{\text{cm}}$ leading to a walk-off between the write pulses and the probe pulses of about $180\,$fs after $5\,$cm of propagation. The resulting reduced refractive index modulation experienced by the probe pulse leads to a reduction in conversion speed and ultimately in conversion efficiency.  On the other hand, there is a zero group delay wavelength at about $\lambda_{\text{p,ideal}}=1550$\,nm, where no walk-off will occur. At these specific probe wavelengths the conversion efficiency exhibits a local minimum and a local maximum, respectively. However, the local conversion efficiency  maximum at $\lambda_{\text{p,ideal}}$ does not correspond to a global one, as the phase mismatch increases approximately linearly with the detuning from the write wavelength $\lambda_{\text{write}}$ and thereby reduces the maximum achievable conversion efficiency. This reduction can be compensated by the coupling strength of the LPG, which is proportional to the peak power of the write beam: in Fig.~\ref{Fig:5}b the conversion efficiency at a probe wavelength of 1550\,nm is shown depending on the write modes' peak power. For low peak powers almost no conversion can occur before write and probe beam are out of phase resulting in the start of back-conversion. With increasing peak power of the write beam the conversion efficiency exceeds 90\% at a peak power of 80kW despite the existing phase mismatch $\Delta(\lambda)$ because of an increased conversion speed. The increased coupling strength also lessens the influence of the chromatic walk-off as the conversion will have taken place before the pulses do not overlap anymore. For peak powers of 120\,kW in the write modes (compare red crosses in Fig.~\ref{Fig:5}a) this leads to a conversion efficiency of more than 90\% up to probe wavelengths of 1700\,nm. However, note that an increase in write peak power also widens the wavelength range wherein direct four-wave mixing occurs and renders mode-conversion undetectable.

The presented results now indicate that a dual-color conversion scenario is feasible using, e.g., Ytterbium- as well as Erbium-doped fiber laser systems for writing an LPG at 1030\,nm and probing it with negligible walk-off at 1550\,nm or vice versa. The converted probe pulse can then very easily be distinguished from the write pulse by spectral filtering as both pulses are separated in center wavelength by over $500$\,nm corresponding to a frequency displacement of about 100\,THz.

\section{All-optical modulation}
\label{sec:Modu}
\begin{figure*}[htbp]
\centerline{\includegraphics[width=2\columnwidth]{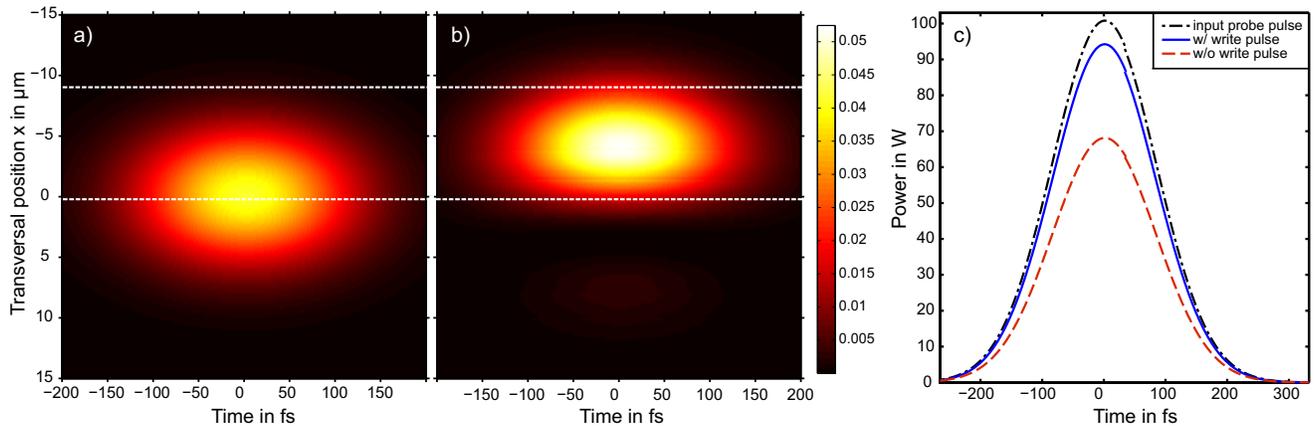}}
\caption{\label{Fig:4} Spatio-temporal profile of the probe pulse ($\lambda=1550$\,nm) reduced to two dimensions by taking a slice through the spatial profile at $y=0\,\mu$m. Depicted is the color-coded intensity of the probe pulse after 31.8\,mm of propagation: a) without the influence of a write pulse  and b) with a co-propagating write pulse at $1030$\,nm present (the \LPx01- and the \LPx11-mode have equal peak power). The dashed white lines mark the position where a single mode fiber is assumed to be spliced to the few-mode fiber with a transverse offset. The resulting time dependent transmission into the fundamental mode of the single mode-fiber with and without write pulse presented is depicted in c).}
\end{figure*}

The results presented above enable various all-optical switching and modulation applications as they allow to control the mode content of the probe beam, and thereby its spatial distribution, with a write beam. In order to demonstrate the quality of the control, the spatio-temporal profile of the probe pulse is investigated in Fig.~\ref{Fig:4}a and b. For this purpose the fiber is considered to be cut off at the point of equal peak powers of both probe modes. To obtain the spatio-temporal output field, the numerically calculated mode fields of all the fiber modes are  summed up after spectral filtering with a pass band of 80\,nm around a center wavelength of $1550$\,nm to get rid of the write modes. In Fig.~\ref{Fig:4}a the probe pulse is solely propagating along the fiber and the Gaussian input profile is preserved in space and time. If a write pulse is now coupled into the fiber, the output probe mode field of the fiber shifts transversely due to the multimode interference between the fundamental \LP01- and the higher-order \LP11-mode (see Fig.~\ref{Fig:4}b). As the Gaussian temporal shape is preserved during conversion for both modes, this shift is uniform over the whole pulse duration. The presented result is thereby a considerable improvement to the results of reference \cite{Walbaum2012} where the mode content varied over the pulse leading to a spatio-temporal inhomogeneous output. The achieved temporal uniform control over the mode content now allows high-quality all-optical modulation, e.g.  by splicing a single-mode fiber with a transverse offset to the multimode fiber resulting in a functionality similar to a saturable absorber.  The position and $1/e^2$-width of such a filtering fiber are indicated by the dashed white lines in Fig.~\ref{Fig:4}a and b. The coupling of the probe beam into the fundamental mode of the single-mode fiber will then depend on the transverse offset induced by the write beam. In Fig.~\ref{Fig:4}c the temporal transmission into the fundamental mode of the filtering fiber is presented with and without an LPG present by calculating the overlap integral for each time slice of the probe pulse with the target mode profile. When the write pulse is present, 93\% of the energy of the probe pulse is transmitted through the single-mode fiber with 7\% being lost due to a non-perfect spatial mode-matching comparable to the non-saturable loss of a saturable absorber. The transmission without an induced LPG drops down to 67\% leading to a modulation depth of about 26\%. It should be noted that even higher modulation depths are possible by using a larger transverse offset of the single-mode fiber, although this comes at the cost of a reduced maximum transmission or non-saturable loss. The pulse energy necessary for the demonstrated modulation scheme is in the order of $75\,$nJ and might thereby hamper an easy application. Utilizing few-mode fibers or waveguides made from highly-nonlinear materials, e.g., made from chalcogenide glasses with an increased nonlinearity of about 2 orders of magnitude compared to fused silica \cite{Eggleton2011}, has the capability to further reduce the requirement of high pulse energies for the write bream in the future. However, the numerically demonstrated performance of this scheme already exceeds the performance of commercially available semiconductor saturable absorbers with respect to modulation depth and non-saturable loss and could therefore at least theoretically be used, e.g., for low loss synchronization of two different laser systems as it has been shown in \cite{Walbaum2010}.

\section{Conclusion}
In this publication we studied extensively by numerical simulations the use of graded-index fibers in order to overcome the limitations previously encountered in step-index fibers concerning the conversion of transverse modes at long-period gratings which are optically induced by femtosecond light pulses. In a scenario similar to earlier work in step-index fibers, it was found that, due to the flat phase-matching curve ($\Delta(\lambda)<0.04\cdot \frac{2\pi}{\text{cm}}$ from $1030\,$nm up to $1800\,$nm) as well as due to the very small group delay difference of $0.16\frac{\text{fs}}{\text{cm}}$ between the first two transverse modes, highly efficient mode conversion with less than 1.5\% residual energy in the fundamental mode is possible. Furthermore, the temporal as well as the spectral shape of the probe pulses are almost preserved during conversion, allowing an optically induced high quality mode control. Finally, a  dual-color approach for the involved light fields was successfully demonstrated in graded-index fibers. The separation of the pump from the probe beam by spectral filtering should allow to circumvent the experimental challenges, that arise from distinguishing probe and write beam through polarization. Mode conversion was found to be limited by the combination of wavelength-dependent walk-off between probe and write pulses as well as phase mismatch. A faster conversion speed through increased coupling strength from higher write beam peak power was presented as means to compensate walk-off as well as phase-mismatch: using write pulses with 120\,kW peak power mode conversion could be demonstrated over a very broad wavelength region with high efficiency ($>90\%$ from 1160\,nm up to 1700\,nm).

%

%
%
%
%

\end{document}